\documentstyle[twocolumn,eclepsf]{jpsj}

\def\runtitle{Frustrated Spin System
in $\theta$-(BEDT-TTF)$_2$RbZn(SCN)$_4$}
\def\runauthor{Hitoshi {\sc Seo} and Hidetoshi Fukuyama}

\def\lsim{\lower -0.3ex \hbox{$<$} \kern -0.75em \lower 0.7ex \hbox{$\sim$}}
\def\gsim{\lower -0.3ex \hbox{$>$} \kern -0.75em \lower 0.7ex \hbox{$\sim$}}

\title{Frustrated Spin System \\
in $\theta$-(BEDT-TTF)$_2$RbZn(SCN)$_4$ }
\author{Hitoshi {\sc Seo}\footnote{E-mail: hseo@watson.phys.s.u-tokyo.ac.jp}
and Hidetoshi Fukuyama}
\inst{Department of Physics, Faculty of Science, University of Tokyo, Tokyo 113}

\recdate{}

\abst{The origin of the spin gap behavior in the low-temperature dimerized 
phase of $\theta$-(BEDT-TTF)$_2$RbZn(SCN)$_4$ has been theoretically studied 
based on the Hartree-Fock approximation for the 
on-site Coulomb interaction at absolute zero. 
Calculations show that, in the parameter region  
considered to be relevant to this compound, antiferromagnetic ordering 
is stabilized between dimers 
consisting of pairs of molecules coupled with the largest transfer integral.  
Based on this result an effective localized spin 1/2 model is constructed 
which indicates the existence of the frustration among spins. 
This frustration may result in the formation of spin gap.}
\kword{BEDT-TTF, spin gap, Hartree-Fock approximation, 
on-site Coulomb interaction, antiferromagnetic spin ordering, 
frustration}

\begin{document}
\sloppy
\maketitle

Two-dimensional (2D) organic conductors composed of 
BEDT-TTF (abbreviated as ET) 
have been attracting spread attention 
due to the variety of their electronic properties. 
In spite of the fact that the electron concentration of the conducting ET layer
is fixed at 3/4 in 
compounds expressed as (ET)$_2X$, where $X$ are anions, 
various phases such as paramagnetic 
metal, superconductor, paramagnetic insulator, 
antiferromagnetic (AF) insulator and even a spin-gapped (SG) insulator emerge. 
To understand these diverse behaviors, 
the role of Coulomb interaction is crucial because 
its value is considered to be of the order of the bandwidths 
in these compounds. 
Such effects of on-site Coulomb interaction 
have been investigated systematically
by Kino and Fukuyama within the Hartree-Fock (HF) approximation
for $\kappa$-(ET)$_2X$, $\alpha$-(ET)$_2$I$_3$ 
and $\alpha$-(ET)$_2M$Hg(SCN)$_4$. \cite{Kino,Kino2}
It was shown by them that the ground state properties of 
$\kappa$-(ET)$_2X$ can be explained by considering 
each dimer as a unit\cite{Kino2} 
as was argued by Kanoda on experimental grounds. \cite{Kanoda}
The term `dimer' denotes the pair of parallel molecules 
between whom the transfer integral 
is noticeably larger than the other transfer integrals.
Hence the AF insulating phase of $\kappa$-(ET)$_2X$ is considered to be 
a Mott insulator due to the existence of 
one hole per dimer.\cite{Kanoda,Miyagawa} 

Such theoretical procedures of combing the HF calculations and 
the quantum spin model so derived also
allow us to understand the possible origin of 
the SG behavior in $\lambda$-(BETS)$_2$Ga$X_zY_{4-z}$,
whose donor is a seleno-analog molecule 
of ET, i.e. BETS (abbreviation for BEDT-TSeF). \cite{Seo}
The existence of a SG in this system had originally been 
suggested from magnetic susceptibility 
measurements,\cite{Koba} and 
its origin was not due to the spin-Peierls (SP) transition. 
This compound has a dimeric structure similar to $\kappa$-(ET)$_2X$; 
the intradimer transfer integral is about twice that of 
the other transfer integrals. 
Thus the insulating phase can be considered as a 2D 
localized spin-1/2 system where each hole is localized on a dimer 
but an intrinsic alternation of the superexchange couplings between 
these spins exists, 
which may lead to the formation of SG.\cite{Seo} 

\begin{figure}
\begin{center}
\epsfile{file= 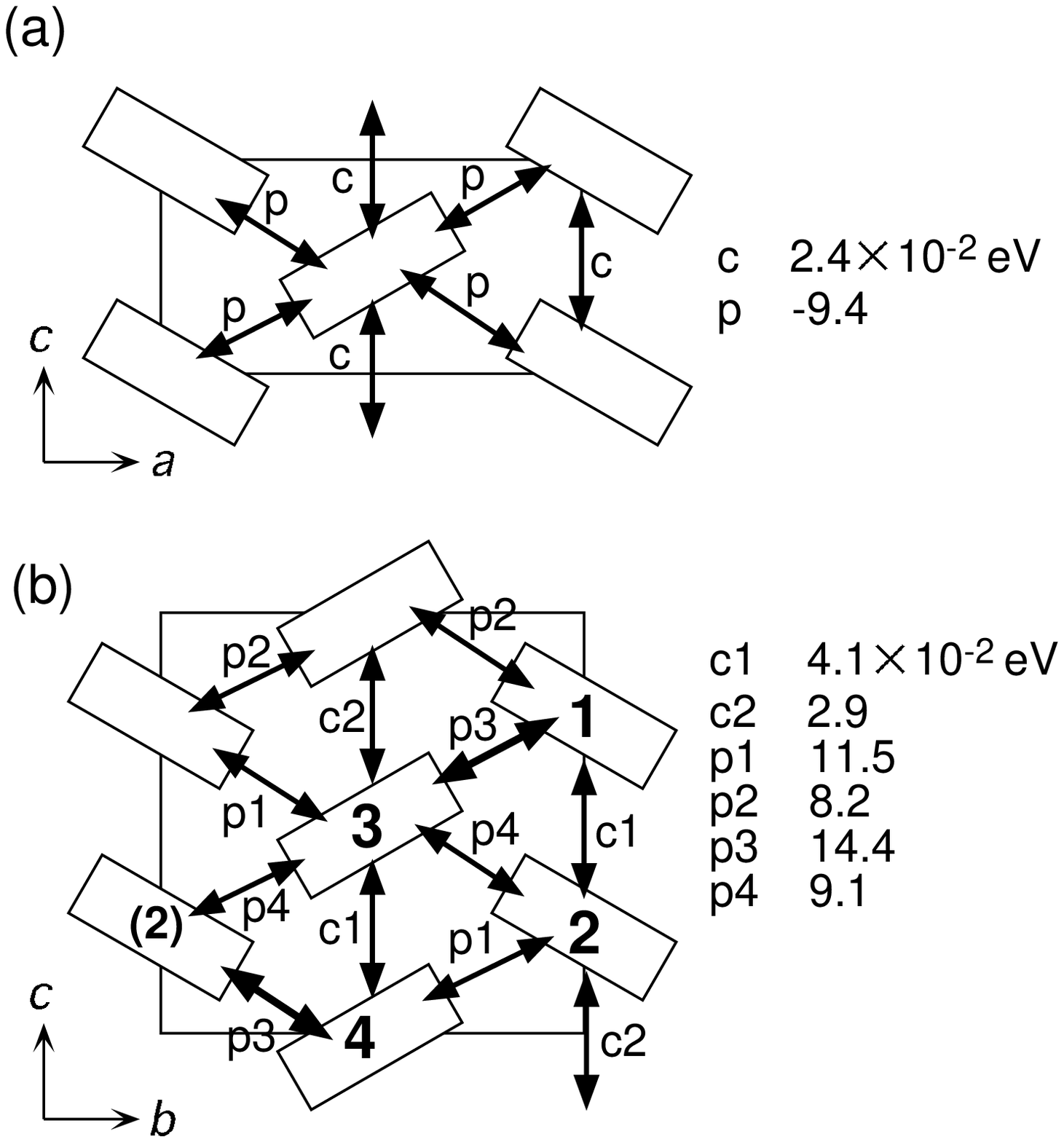,height=9.5cm}
\end{center}
\caption{Schematic donor structure of $\theta$-(BEDT-TTF)$_2$RbZn(SCN)$_4$ 
in the high-temperature phase (a) and 
the slowly cooled low-temperature phase (b). 
Each ET molecule is schematically expressed as a rectangle. 
The values of transfer integrals 
are given by $t=Es$, where $E=-10$ eV and 
$s$ is the value of the overlap integral 
taken from ref. 
9.}
\label{structure}
\end{figure}

Recently another 2D organic compound $\theta$-(ET)$_2$RbZn(SCN)$_4$ 
has been synthesized which shows a SG behavior. \cite{Mori}
The existence of a SG has been suggested from magnetic 
susceptibility\cite{Mori2,Mori3} and NMR\cite{Nakamura} measurements. 
The $\theta$-type ET compounds in general have a donor structure 
with two ET molecules in a unit cell
as shown in Fig. \ref{structure}(a) and have a simple 3/4-filled band, 
that is, they do not have a dimeric structure. 
However, in the case of $\theta$-(ET)$_2$RbZn(SCN)$_4$, 
when it is slowly cooled,  a metal-insulator transition occurs at 190 K 
accompanied by a lattice modulation
\cite{Mori3}
which leads to the dimerization along the $c$-direction, 
thus the unit cell volume doubles and the transfer 
integrals become modulated as in Fig. \ref{structure}(b). 
No SP lattice 
distortion has been observed so far\cite{Mori3} in the temperature range 
where the magnetic susceptibility shows a steep decrease. 
It can be seen in Fig. \ref{structure}(b)  
that molecule pairs (1-3) and (2-4) 
with intradimer transfer integrals $t_{p3}$
may be considered as dimers, 
but the value of $t_{p3}$ is not so large as compared to 
the other transfer integrals. 
Since in organic conductors such as (TMTSF)$_2X$ and $\kappa$-(ET)$_2X$ 
the dimers consist of parallel molecules, 
we may even be able to consider 
the parallel (1-2) and (3-4) molecule pairs as dimers 
possessing an intradimer transfer integral $t_{c1}$.  

The aim of this study is to elucidate whether dimers can be considered 
as units in the donor layer, and if it is so, 
which pairs of molecules should be considered as dimers. 
Furthermore, based on these results, 
the origin of the SG phenomenon will be investigated. 

In order to theoretically study the electronic structure of the slowly cooled 
low temperature phase of $\theta$-(ET)$_2$RbZn(SCN)$_4$, 
only its 2D donor plane
(Fig. \ref{structure}(b)) is considered where the 
transfer integrals between ET molecules are assumed as in 
Fig. \ref{structure}(b) together with 
the on-site Coulomb interaction $U$ on ET molecule. 
The Hamiltonian is given as follows; 
\begin{eqnarray}
H=\sum_{<i,j>}\sum_{\sigma} \left( t_{i,j}a^{\dagger}_{i\sigma}%
    a_{j\sigma}+h.c.\right) +
    U\sum_{i}n_{i\uparrow}n_{i\downarrow},
\label{eqn:Hamil}
\end{eqnarray}
where $<i,j>$ denotes the neighboring site pair, 
$\sigma$ is the spin index which takes $\uparrow$ and $\downarrow$, 
$n_{i\sigma}$ and  $a^{\dagger}_{i\sigma}$ ($a_{i\sigma}$) denote 
the number operator and the creation (annihilation) operator for the 
electron of spin $\sigma$ at the $i$th site, respectively. 

\begin{figure}
\begin{center}
\epsfile{file= 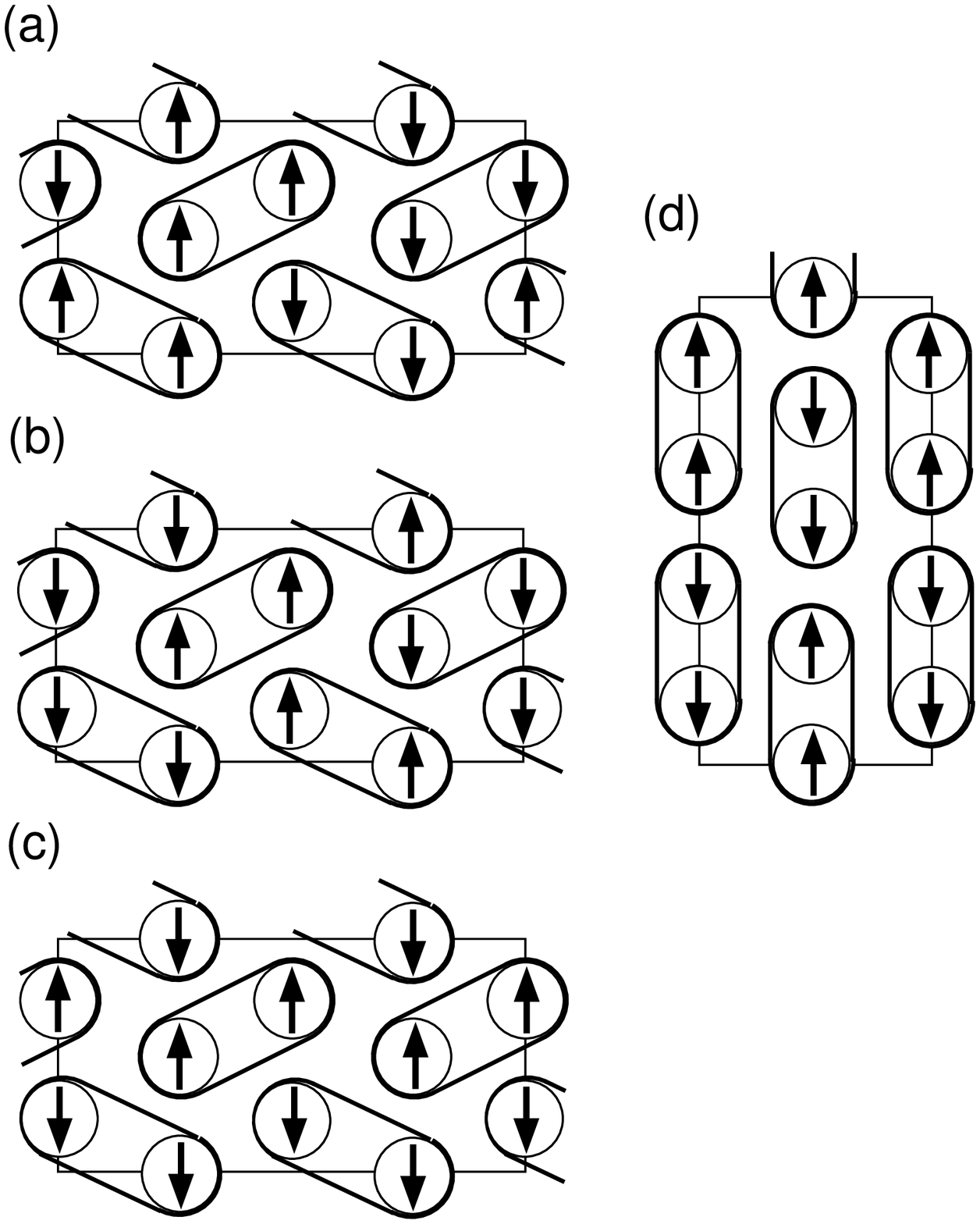,height=9cm}
\end{center}
\caption{Different types of AF ordering for the structure given in 
Fig. \protect\ref{structure}. Circles and ellipses 
represent ET molecules and dimers, respectively. 
The magnetic unit cells are shown by rectangles.}
\label{AF}
\end{figure}

The Coulomb interaction $U$ is treated in the HF approximation 
in a manner similar to that in refs. 1 and 2.
Our calculations are carried out at $T=0$ and 
the average electron density is fixed 
at 3/4-filling, namely one hole per two ET molecules as in the real compound. 
Several types of AF solutions are taken into account as shown in 
Fig. \ref{AF} and 
their energies are compared as a function of $U$
so that the true ground state is searched for. 
The total energy ${\cal E}$ is calculated as 
\begin{equation}
{\cal E}=\frac 1{N_{cell}}\sum_{lk\sigma}\epsilon_{lk\sigma}
n_F\left(\epsilon_{lk\sigma}\right)
-U\sum_i\left<n_{i\uparrow}\right>\left<n_{i\downarrow}\right>,
\end{equation}
where $N_{cell}$ is the total number of cells, 
$\epsilon_{lk\sigma}$ is the $l$th eigenvalue at each $k$ and 
$n_F\left(\epsilon \right)$ is the Fermi distribution function. 
Due to the structure of interdimer transfer integrals which we will
discuss later, 3 different types of AF (a),(b) and (c), 
as shown in Fig. \ref{AF} exist, 
where the spins within the pairs (1-3) and (2-4) 
are parallel and AF ordering occurs between the pairs.  
In the case of (d) type AF ordering, 
the spins within the (1-2) and (3-4) pairs are parallel 
and 2D AF ordering takes place between pairs. 

\begin{figure}
\begin{center}
\epsfile{file= 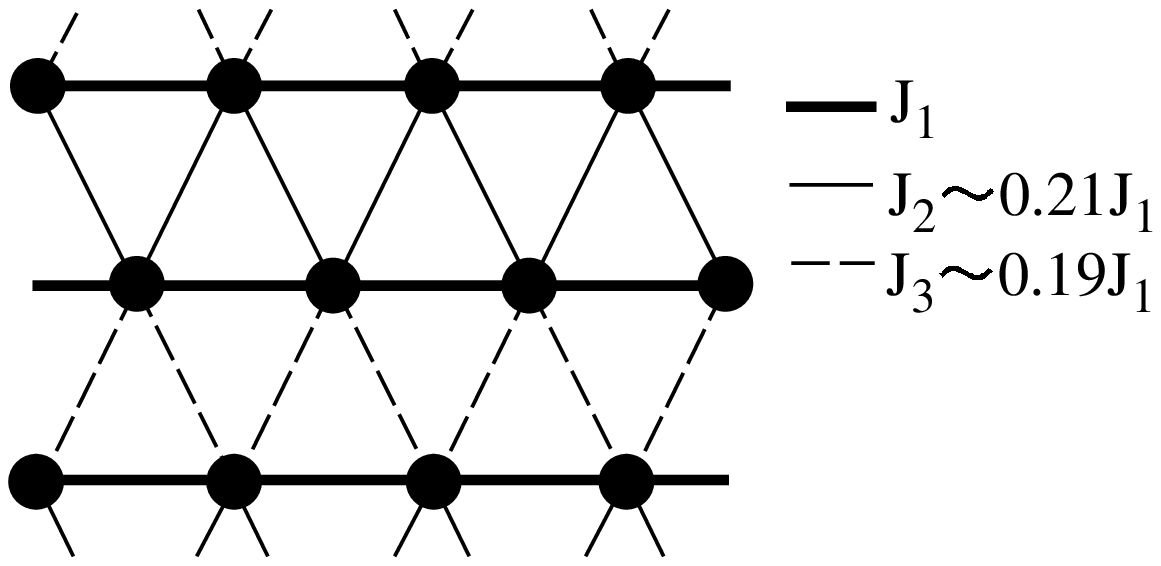,height= 6cm}
\end{center}
\caption{Energy of each state, $E_{\alpha}$, shown in Fig. \protect\ref{AF} 
relative to that of the paramagnetic solution, $E_p$, 
$\Delta E(\alpha) = E_{\alpha}-E_p$. }
\label{energy}
\end{figure}

The calculated energies of each state, $\alpha$, relative to the paramagnetic 
solution, $\Delta E(\alpha) = E_{\alpha}-E_p$, 
for four different types of AF solutions 
are shown in Fig. \ref{energy}.
These results show that (a) and (b) types of AF, which are degenerate, 
are stabilized 
within realistic values of $U$, namely $U=0.6-1.0$ eV. 
In these (a) and (b) types AF, 
the hole density on each ET molecule is similar 
but differs by a few percent from +0.5, 
and the magnitude of spin moments is 
$0.8-0.9 \mu_B$ per dimer. 
Hence we can consider that 
the electrons are almost fully localized on each dimer. 

\begin{figure}
\begin{center}
\epsfile{file= 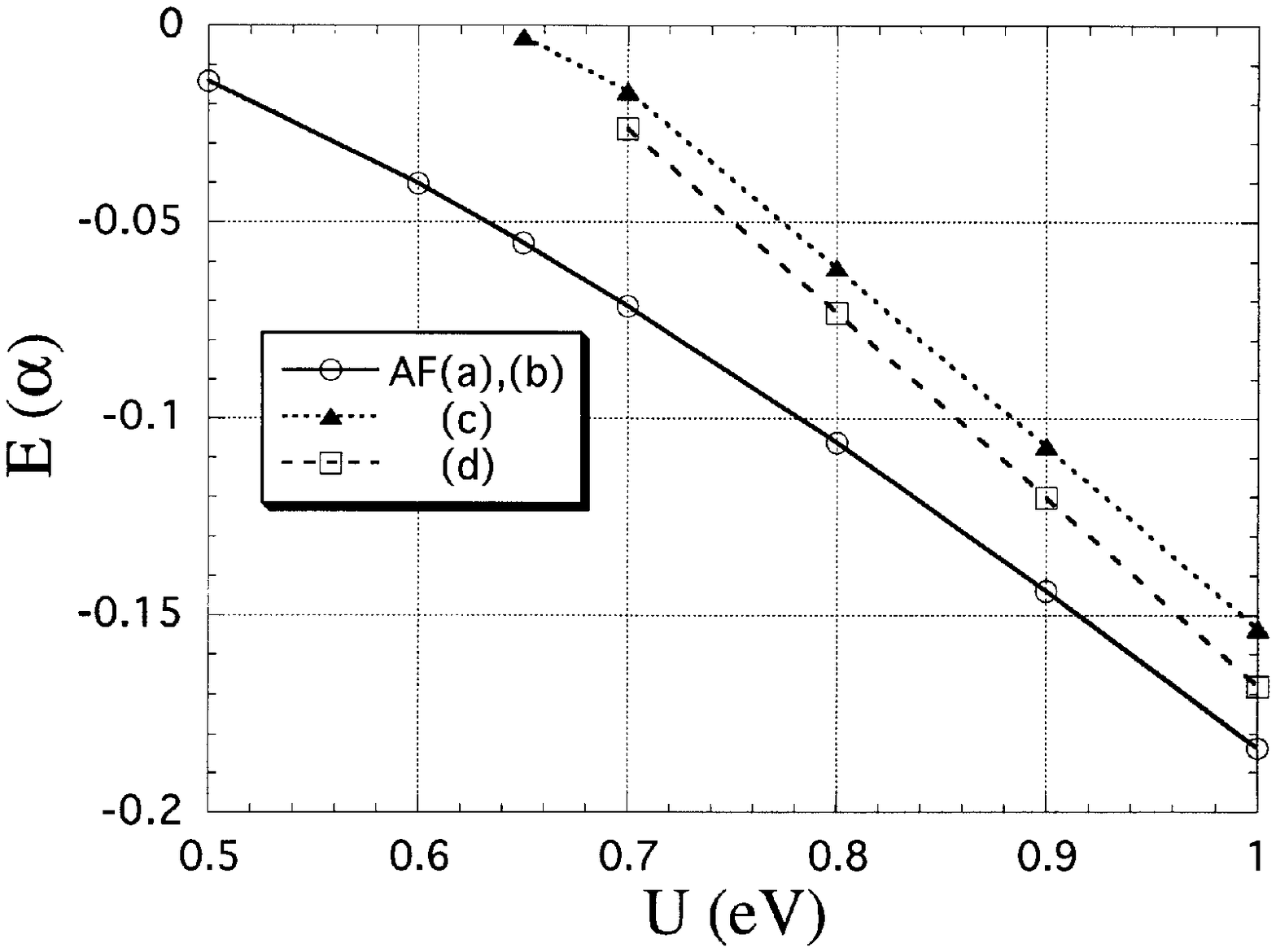,height= 4cm}
\end{center}
\caption{Effective dimer model for $\theta$-(ET)$_2$RbZn(SCN)$_4$. 
Full circles represent dimers and the values of the effective 
superexchange couplings are shown. 
Thick, thin and thin-dotted lines represent the spin exchange couplings, 
$J_1, J_2$ and $J_3$, respectively. }
\label{dimer}
\end{figure}

From the above results, we shall consider the pairs (1-3) and (2-4) as dimers. 
As shown in refs. 2 and 5 
the resulting dimer model can be constructed 
by estimating the superexchange couplings 
using the relation  $J\simeq 4t^2/U_{dimer}$
($t$ is the effective interdimer transfer energy and 
$U_{dimer}$ is the effective Coulomb interaction 
between two holes in a dimer), 
which is given in Fig. \ref{dimer} with the magnitudes of 
$J_2\simeq 0.21J_1$ and $J_3\simeq 0.19J_1$. 
This estimate shows that $J_1$ is much larger than $J_2$ and $J_3$
is in accordance with 
the HF results where (a) or (b) type AF is stabilized. 

\begin{figure}
\begin{center}
\epsfile{file= 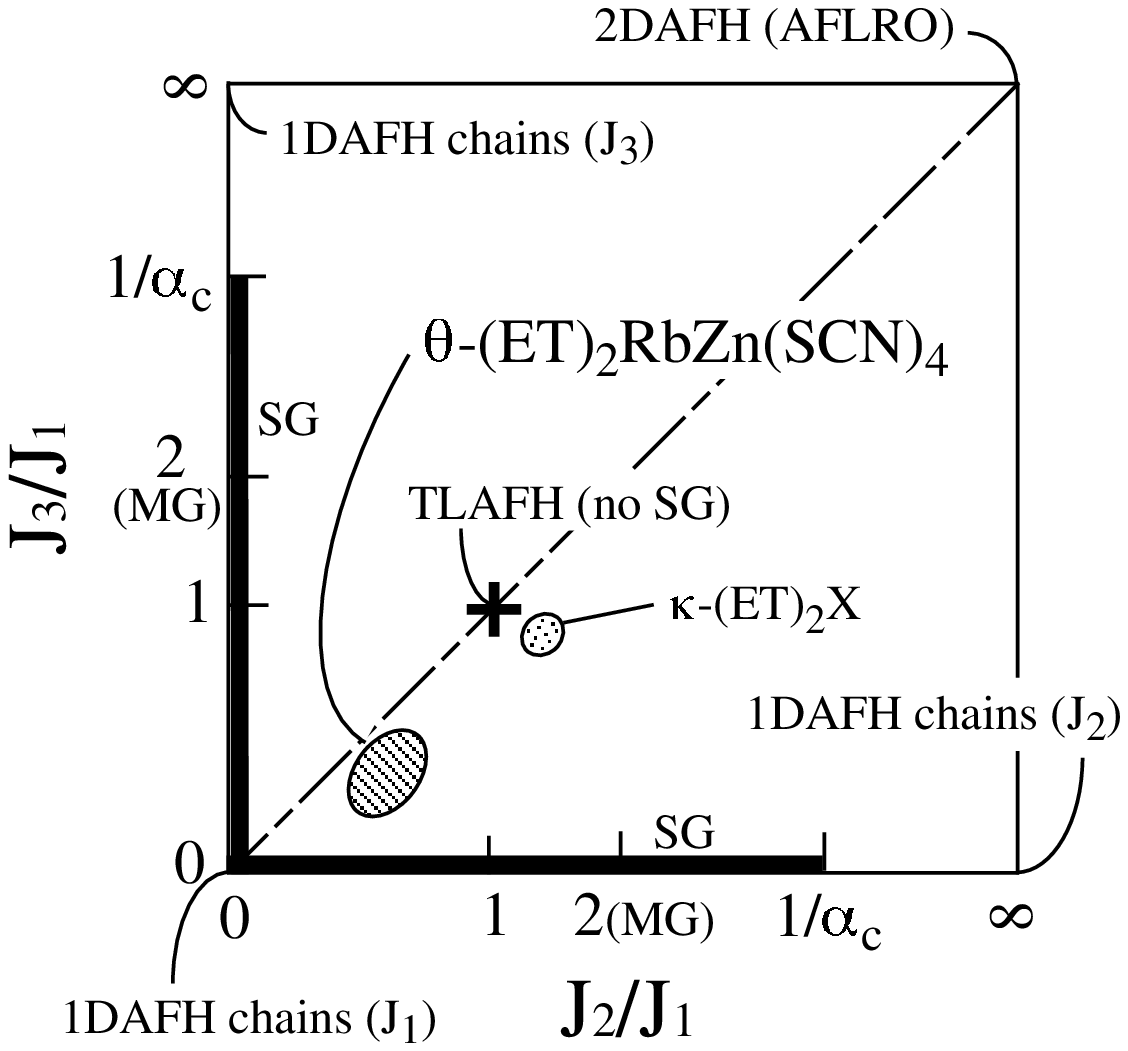,height= 8cm}
\end{center}
\caption{The schematic phase diagram of the Heisenberg spin model 
described in Fig. \protect\ref{dimer}. \protect\cite{Katoh} 
It is symmetric with respect to the dot-dashed line. 
SG, $n$DAFH, TLAFH, AFLRO and MG 
represent spin gap, $n$-dimensional antiferromagnetic Heisenberg model, 
triangular lattice antiferromagnetic Heisenberg model, 
antiferromagnetic long range order and 
Majumdar-Ghosh point, respectively. 
The regions where a spin gap is known to exist are shown by the thick lines. 
The regions where $\theta$-(ET)$_2$RbZn(SCN)$_4$ and $\kappa$-(ET)$_2$X
are expected to be located  
are shown by the shaded and the dotted areas, respectively.}
\label{Katoh}
\end{figure}

The quantum fluctuations among spins are taken into account in this model. 
The phase diagram of the ground state of 
such a 2D spin-1/2 model on the plane of $J_2/J_1$ and $J_3/J_1$, 
can be deduced as in Fig. \ref{Katoh}. \cite{Katoh}
When $J_3=0$ ($J_2=0$), the model is identical to 
spin 1/2 chains coupled by nearest-neighbor interaction $J_2$ ($J_3$) 
and next-nearest-neighbor interaction $J_1$ which 
induces frustration between spins. 
In this model 
it is known\cite{1DJ1J2} that 
when the ratio of the nearest-neighbor and the 
next-nearest-neighbor interaction, $J_2/J_1$ for $J_3=0$, for example, 
is increased, 
the low energy exitation changes its behavior 
from the one with a finite gap like the Majumdar-Ghosh model ($J_2/J_1=2$) 
to a gapless one like a Heisenberg chain 
at $J_2/J_1 = 1/\alpha_c; \alpha_c=0.2411$\cite{Okamoto} 
(see the axis in Fig. \ref{Katoh}). 
Similar behavior exists on increasing $J_3/J_1$ at $J_2=0$. 
When $J_2=J_3$ (the dot-dashed line in Fig. \ref{Katoh}), 
the model reduces to a 2D AF Heisenberg model 
with nearest-neighbor coupling $J_2=J_3$ and 
next-nearest-neighbor coupling $J_1$ along 
{\em only one} diagonal direction. \cite{Horsh}
Its ground state 
is unknown except for the case of $J_1=J_2=J_3$, which is the simple 
triangular lattice with no SG\cite{sankaku}. 
This is approximately similar to the case of  
the dimer model for $\kappa$-(ET)$_2$X\cite{Kino,Kino2}
as shown in Fig. \ref{Katoh} and 
the ground state experimentally observed there 
is actually an AF insulator. \cite{Miyagawa}
(2D Heisenberg model with nearest-neighbor and next-nearest-neighbor 
along {\em both} diagonal directions has been well 
studied. \cite{2DJ1J2}) 
As seen in Fig. \ref{Katoh}, $\theta$-(ET)$_2$RbZn(SCN)$_4$ 
is located in the unexplored region, 
yet in this region frustration is large such that the ground state 
may have a SG. 
This is the possible explanation we propose in this paper. 

Here we note that if $\theta$-(ET)$_2$RbZn(SCN)$_4$ is in the region of AF, 
the SP transition will be a candidate for the origin of the 
SG behavior because $J_1$ is much larger than 
$J_2$ and $J_3$. It is known that the effect of interchain coupling 
on SP transition is to enhance the tendency toward 
AF long range order. \cite{Inagaki} 
However, the competition between SP transition 
and N\'{e}el ordering 
in the presence of frustration among the interchain couplings 
is also an interesting unanswered question. 

In SG systems, the impurity effect is interesting 
because impurities easily destroy the SG and 
induce AF long range order. \cite{SG+AF}
If some kind of disorder is introduced into 
$\theta$-(ET)$_2$RbZn(SCN)$_4$ without any structural changes, 
the HF calculations predict that 
the configuration of the magnetic ordering
will be either (a) or (b) type AF. 
In recent experiments in this direction, however,  
the dimeric structure was unfortunately not maintained; 
in $\theta$-(ET)$_{2-x}$(BMDT-TTF)$_x$RbZn(SCN)$_4$ and 
$\theta$-(ET)$_2$Rb$_{1-x}$Cs$_x$Zn(SCN)$_4$, 
the structural phase transition observed in the pure compound is supressed 
and then the SG collapses. \cite{Nakamura2,Mori4}

In summary, the origin of the SG behavior in the slowly cooled 
dimerized phase in $\theta$-(ET)$_2$RbZn(SCN)$_4$ is studied 
both by the HF calculations and the resulting 
effective dimer spin model. 
HF calculations indicate that the pairs of molecules 
coupled with the largest transfer integrals should be 
considered as dimers, and that the electrons are localized 
on each dimer due to the Coulomb interaction forming a Mott insulator. 
Quantum effect will create frustration among the localized 
spins which is the possible origin of the SG. 

\acknowledgements
The authors thank N. Katoh and H. Kino for useful discussions and suggestions. 
They also thank H. Mori, K. Kanoda, T. Nakamura, A. Kobayashi and T. Takahashi 
for informative discussions from the experimental viewpoint.  
This work was financially supported by a 
Grant-in-Aid for Scientific Research on Priority Area ``Anomalous Metallic
State near the Mott Transition'' (07237102) from the Ministry of Education, 
Science, Sports and Culture.


\end{document}